\documentclass[onecolumn,amsmath,amssymb,pre]{revtex4-1}

\usepackage{graphicx}
\usepackage{color}
\usepackage{dcolumn}
\usepackage{amsmath}
\usepackage{subfigure}

\usepackage{sidecap}
\usepackage{dsfont}

\usepackage{mathrsfs}
\usepackage{amsfonts}
\usepackage{indentfirst}
\usepackage{bm}

\usepackage[colorlinks,citecolor=blue]{hyperref}

\newcommand{\be}{\begin{equation}}
\newcommand{\ee}{\end{equation}}
\newcommand{\bey}{\begin{eqnarray}}
\newcommand{\eey}{\end{eqnarray}}
\newcommand{\bw}{\begin{widetext}}
\newcommand{\ew}{\end{widetext}}

\newcommand{\ra}{\rangle}
\newcommand{\la}{\langle}

\newcommand{\ba}{\begin{array}}
\newcommand{\ea}{\end{array}}
\newcommand{\bi}{\begin{itemize}}
\newcommand{\ei}{\end{itemize}}
\newcommand{\bem}{\begin{enumerate}}
\newcommand{\eem}{\end{enumerate}}

\begin{document}

\title{Mixed eigenstates in the Dicke model: 
Statistics and power-law decay of the relative proportion in the semiclassical limit}

\author{Qian Wang}
\affiliation{CAMTP-Center for Applied Mathematics and Theoretical Physics, University of Maribor, 
Mladinska 3, SI-2000 Maribor, Slovenia, European Union, and \\
Department of Physics, Zhejiang Normal University, Jinhua 321004, China}

\author{Marko Robnik}
\affiliation{CAMTP-Center for Applied Mathematics and Theoretical Physics, University of Maribor, 
Mladinska 3, SI-2000 Maribor, Slovenia, European Union}

\begin{abstract}

How the mixed eigenstates vary with approaching the semiclassical 
limit in mixed-type many-body quantum systems is an interesting but still less known question. 
Here, we address this question in the Dicke model, a celebrated many-body model that
has a well defined semiclassical limit and undergoes a transition to chaos in both
quantum and classical case.  
Using the Husimi function, we show that the eigenstates of 
the Dicke model with mixed-type classical phase space 
can be classified into different types.
To quantitatively characterize the types of eigenstates, we     
study the phase space overlap index, which is defined in terms of Husimi function.
We look at the probability distribution of the phase space overlap index and 
investigate how it changes with increasing system size, 
that is, when approaching the semiclassical limit.
We show that increasing the system size gives rise 
to a power-law decay in the behavior of 
the relative proportion of mixed eigenstates.
Our findings shed more light on the properties of eigenstates in mixed-type
many-body systems and suggest that the principle of uniform semiclassical condensation 
of Husimi functions should also be valid for many-body quantum systems.  

\end{abstract}

\date{\today}

\maketitle

\section{Introduction}

Characterizing and understanding quantum chaos in many-body quantum systems through
the properties of eigenstates is a widely studied topic.
Even though it has been explored since the beginning of quantum chaos, 
the close connection between chaos and the questions 
that arise in recent experimental and theoretical studies has led to a revival in
studying various features of chaotic eigenstates, 
such as their entanglement entropy \cite{Vidmar2017,Murthy2019,Haque2022},  
delocalization or localization \cite{Batistic2013,Batistic2019,Batistic2020,QianW2023a,QianW2020,Villasenor2021}, 
scarring \cite{Cameo2021a,Cameo2022,Chandran2023,Evrard2023}
and multifractality \cite{Backer2019,Pausch2021,Magnani2023} as well. 
Unlike the strongly chaotic systems, the properties of eigenstates in the mixed-type many-body 
systems that show a mixture of regular and chaotic behaviors 
are less commonly explored \cite{Kourehpaz2022,Nakerst2023},
despite the fact that mixed-type behavior is more generic.

In studying the single particle systems, it has been demonstrated that 
the classical counterparts of the mixed-type quantum systems are generally 
associated with mixed-type phase space with coexistence of regular and chaotic motions. 
Based on this fact, Percival was the first to suggest that the eigenstates 
in mixed-type quantum systems can be divided into the chaotic and regular types \cite{Percival1973}.  
After further developments performed by Berry \cite{Berry1977a,Berry1977b}, 
this suggestion becomes the principle of uniform semiclassical condensation 
of Wigner (or Husimi) functions (PUSC) \cite{Robnik1998,Robnik2019,Robnik2020}. 
In the ultimate semiclassical limit, various characters \cite{Prosen1993a,Veble1999}, 
in particular the spectral statistics \cite{Berry1984,Prosen1994b,BaoWen1994,BaoWen1995,Prosen1999,Manos2013}, 
of mixed-type quantum systems can be well understood by means of the PUSC.
Moreover, the correctness of the PUSC for the cases that are not in the asymptotic semiclassical limit 
is also verified in recent works \cite{Lozej2022,QianW2023b}, which are
focused on how the mixed eigenstates evolve as the semiclassical limit is approached. 
However, the situation for the many-body quantum systems is more 
complicated and has not yet been explored.  

In this work, we will investigate whether and how the eigenstates of a mixed-type 
many-body quantum system separate into the regular and chaotic states 
with approaching the semiclassical limit.  
This requires that the considered many body system should have a well defined semiclassical limit.
The model we focus on is the celebrated Dicke model \cite{Dicke1954}. 
An advantage of the Dicke model is that it  
attains its semiclassical limit with increasing the system size $N$ and, thus, allows us to define 
an effective Planck constant $\hbar_{\mathrm{eff}}\simeq1/N$.
A rich variety of phases observed in the Dicke model has made 
it a very popular model to study different phase transitions, 
including thermal phase transition \cite{Hepp1973a,Hepp1973b,Kirton2019,Perez2017,Pragna2022}, 
nonequilibrium phase transitions \cite{Bastidas2012},
quantum phase transition \cite{Nagy2010,Baumann2010,Baumann2011,Romera2012,Emary2003a,Emary2003b}, 
as well as excited state quantum phase transition 
\cite{Perez2017,Perez2011,Magnani2014a,Lobez2016,Brandes2013,Puebla2014,Corps2021}.  
In particular, both quantum and classical Dicke models allow the transition 
from integrability to chaos by changing certain control parameters 
\cite{Emary2003b,Magnani2014b,Magnani2015,Carlos2016}.
As a consequence, it has been widely employed as a paradigmatic model for studying 
quantum chaos 
\cite{Magnani2015,Carlos2016,Magnani2014b,SongL2009,Bhattacharya2014,Lerma2019,Carlos2019} 
and thermalization \cite{Altland2012a,Altland2012b,Swan2019,Lobez2021,Cameo2021b,Villasenor2023}.
It is, therefore, a particularly suitable many-body quantum model for our purpose, as 
a continuation of our previous works \cite{QianW2020,QianW2023b}. 

By carrying out a detailed analysis of the dependence of the degree of chaos on the control parameters for both
classical and quantum models, we determine our studied parameter regions. 
We demonstrate how to identify the types of the eigenstates by comparing their Husimi functions with
the classical Poincar\'{e} section.  
To quantitatively describe different types of eigenstates, 
we utilize the phase space overlap index, which
is defined in terms of eigenstates' Husimi functions 
and has been proved as a valuable tool to quantify eigenstates properties.
We show that the statistics of the phase space overlap index is 
approximately captured by double peak distribution 
with two peaks close to its extreme values,
corresponding to entirely regular and chaotic eigenstates.
Moreover, we find that the relative proportion of the 
mixed eigenstates follows the power-law decay as the semiclassical limit 
is approached, by increasing the system size.
These results are consistent with the ones revealed in the single particle systems,
suggesting the validity of the PUSC for many-body quantum systems. 

The rest of the article is structured as follows. 
In Sec.~\ref{Second}, we introduce the Dicke model. 
Sec.~\ref{SecA} is devoted to the analysis of the semiclassical Dicke model and shows
how chaos develops with varying either the control parameter or the energy of the system.
In Sec.~\ref{SecB}, we discuss the emergence of chaos in the 
quantum Dicke model by means of the statistics of level spacing ratio.
Sec.~\ref{Third} reports our main results of this work.
In Sec.~\ref{TrA}, the Husimi function of an individual eigenstate is defined 
and employed to illustrate different types of the eigenstates. 
The definition of the phase space overlap index and its distribution, 
as well as the variation of the relative proportion of
mixed eigenstates with system size are analyzed in Sec.~\ref{TrB}. 
We finally summarize our findings and conclude in Sec.~\ref{Fourth}.

\section{Dicke model} \label{Second}

As the simplest atom-field system, the Dicke model \cite{Dicke1954} consists of an ensemble of $N$ 
spin-$1/2$ atoms interacting with a single electromagnetic mode within a cavity, 
and its Hamiltonian reads \cite{Emary2003b} (setting $\hbar\equiv1$)
\be \label{DickeH}
   H=\omega a^\dag a+\omega_0 J_z+\frac{2\lambda}{\sqrt{N}}J_x(a^\dag+a),
\ee
where $\omega$ is the frequency of the cavity mode, $\omega_0$ denotes 
the energy splitting of atoms, and $\lambda$ represents the atom-cavity coupling strength.
Here, $a$ ($a^\dag$) is the usual bosonic annihilation (creation) operator, while 
$\mathbf{J}=(J_x, J_y, J_z)$ are the collective pseudospin operators describing $N$ spin-$1/2$ atoms.  
These pseudospin operators fulfill the $\mathrm{SU}(2)$ commutation relations.

The Hamiltonian (\ref{DickeH}) conserves the total spin operator $\mathbf{J}^2$, 
so that the Hilbert space can be separated into different subspaces 
according to the eigenvalues of $\mathbf{J}^2$.
Our study will restrict to the subspace with $j=N/2$ and the 
Hilbert space dimension $\mathcal{D}_\mathcal{H}=(N+1)(\mathcal{N}_{trc}+1)$.
Here, $\mathcal{N}_{trc}$ denotes the truncation number of the bosonic basis.  
Moreover, since the parity operator $\Pi=e^{i\pi(a^\dag a+J_z+j)}$ is also 
a conserved quantity, $[H,\Pi]=0$, one can further divide the Hilbert 
space into even- and odd-parity blocks.
In this work, we focus on the even-parity sector with even $j$. 
As a result, the dimension of the Hilbert space is 
$\mathcal{D}_{\mathcal{H}}^e=(N/2+1)(\mathcal{N}_{trc}+1)-\mathcal{N}_{trc}/2$.
To guarantee the convergence of the numerical results, the value of 
$\mathcal{N}_{trc}$ should be sufficiently large.
We have carefully checked that the results obtained in this work are 
converged for our chosen values of $\mathcal{N}_{trc}$, whose maximal value was $300$.

 \begin{figure}
  \includegraphics[width=\columnwidth]{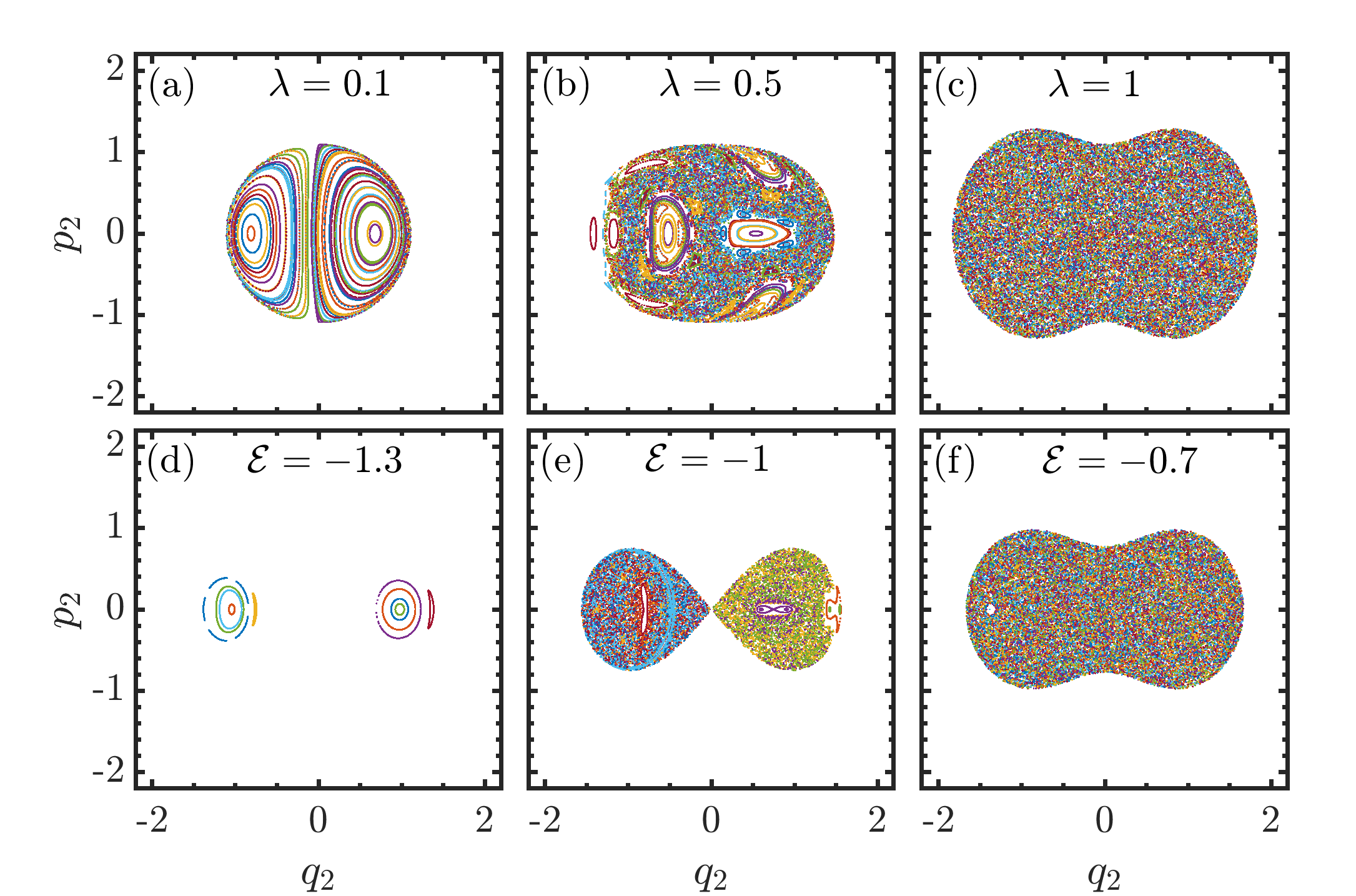}
  \caption{Classical Poincar\'{e} sections of the system (\ref{ClassicalH}) in $(q_2,p_2)$ plane.
  (a)-(c) show the Poincar\'{e} sections for several coupling strengths with fixed energy $\mathcal{E}=-0.4$, 
  while (d)-(f) are the Poincar\'{e} sections for different energies with $\lambda=0.8$. 
  Other parameters: $\omega=\omega_0=1$.
  All quantities are dimensionless.}
  \label{CPS}
 \end{figure}
 
 It was known that the ground state of the model undergoes a second-order quantum phase transition 
 at the critical point $\lambda_c=\sqrt{\omega\omega_0}/2$, which separates
 the normal phase with $\lambda<\lambda_c$ from the 
 superradiant phase with $\lambda>\lambda_c$ \cite{Emary2003b}. 
 Apart from the ground state quantum phase transition, the excited energy spectrum of the system also
 exhibits an excited state quantum phase transition \cite{Caprio2008,Stransky2014,Cejnar2021} 
 into the superradiant phase \cite{Lobez2016,Corps2021,Magnani2014a,
 Brandes2013,Perez2017,Puebla2014,Puebla2013}.  
 In particular, both ground and excited state quantum phase transitions are associated with 
 the onset of classical and corresponding quantum chaos 
 \cite{Emary2003a,Emary2003b,Perez2011,Magnani2014a,Lobez2016}.

\subsection{Chaos in semiclassical system} \label{SecA}

In the semiclassical limit with $N\to\infty$, the Hamiltonian (\ref{DickeH}) turns into its classical counterpart.
To obtain the effective classical Hamiltonian, we utilize 
the Glauber and Bloch coherent states \cite{Radcliffe1971,ZhangW1990}, 
which are, respectively, defined as
\begin{align}
  &|\alpha\ra=e^{|\alpha|^2/2}e^{\alpha a^\dag}|0\ra, \notag \\
  &|\xi\ra=\frac{1}{(1+|\xi|^2)^j}e^{\xi J_+}|j,-j\ra.
\end{align}
Here, $\alpha, \xi$ are the complex parameters, $|0\ra$ is the bosonic field vacuum, 
and $|j,-j\ra$ is the ground state of the atomic sector. 
Then, according to the properties of the coherent states, it is straightforward to find the following relations:
\begin{align}
  &\la\alpha|a^\dag a|\alpha\ra=|\alpha|^2, \notag \\
  &\la\xi|J_x|\xi\ra=j\left(\frac{\xi+\xi^\ast}{1+|\xi|^2}\right), \notag \\
  &\la\xi|J_z|\xi\ra=-j\left(\frac{1-|\xi|^2}{1+|\xi|^2}\right).
\end{align}
The classical Hamiltonian is the expectation of the quantum Hamiltonian (\ref{DickeH}) 
in the tensor product of the coherent states, $|\mathrm{CS}\ra=|\alpha\ra\otimes|\xi\ra$ 
\cite{Carlos2016,Aguiar1992,Bakemeier2013}.
Using above relations, one can easily find that the classical Hamiltonian can be written as
\be \label{CSH}
  \mathcal{H}_c=\frac{\la\mathrm{CS}|H|\mathrm{CS}\ra}{j}
    =\omega\frac{|\alpha|^2}{j}-\omega_0\left(\frac{1-|\xi|^2}{1+|\xi|^2}\right)
    +\frac{2\lambda}{\sqrt{N}}(\alpha^\ast+\alpha)\left(\frac{\xi+\xi^\ast}{1+|\xi|^2}\right).
\ee

To express $\mathcal{H}_c$ in terms of the classical canonical variables, 
we parametrize $\alpha$ and $\xi$ as follows \cite{Corps2021,Cameo2021a,Aguiar1992}
\be
  \alpha=\sqrt{\frac{j}{2}}(q_1+ip_1),\ \xi=\frac{q_2+ip_2}{\sqrt{4-p_2^2-q_2^2}},
\ee
where $(p_1,q_1)\in\mathbb{R}^2$ are the canonical variables for the bosonic sector, while
$(p_2,q_2)\in\mathbb{R}^2$ and $p_2^2+q_2^2\leq4$ denote the 
atomic sector canonical variables \cite{Footnote}. 
Finally, inserting the parametrized $\alpha$ and $\xi$ into Eq.~(\ref{CSH}), 
after some algebra, the classical Hamiltonian is given by
\be \label{ClassicalH}
  \mathcal{H}_c=\frac{\omega}{2}(p_1^2+q_1^2)+\frac{\omega_0}{2}(p_2^2+q_2^2)
     +\lambda q_1q_2\sqrt{4-p_2^2-q_2^2}-\omega_0.
\ee
The classical system is described by the canonical variables 
$\mathbf{x}=(q_1,q_2,p_1,p_2)\in\mathbb{R}^4$, meaning that the classical phase space is four dimensional. 
From the classical Hamiltonian (\ref{ClassicalH}), the classical equations of motion are given by
\begin{align}\label{CEM} 
   &\dot{q_1}=\frac{\partial H_c}{\partial p_1}=\omega p_1,\quad 
     \dot{q_2}=\frac{\partial\mathcal{H}_c}{\partial p_2}
      =\omega_0 p_2-\frac{\lambda q_1q_2p_2}{\sqrt{4-p_2^2-q_2^2}},\notag \\
   &\dot{p_1}=-\frac{\partial\mathcal{H}_c}{\partial q_1}=-\omega q_1-\lambda q_2\sqrt{4-p_2^2-q_2^2}, \notag \\
   &\dot{p_2}=-\frac{\partial\mathcal{H}_c}{\partial q_2}
     =-\omega_0 q_2-\lambda q_1\sqrt{4-p_2^2-q_2^2}+\frac{\lambda q_1q_2^2}{\sqrt{4-p_2^2-q_2^2}}, 
\end{align}
associated with the initial condition $\mathbf{x}_0=(q_{1,0},q_{2,0},p_{1,0},p_{2,0})$.

 \begin{figure}
  \includegraphics[width=\columnwidth]{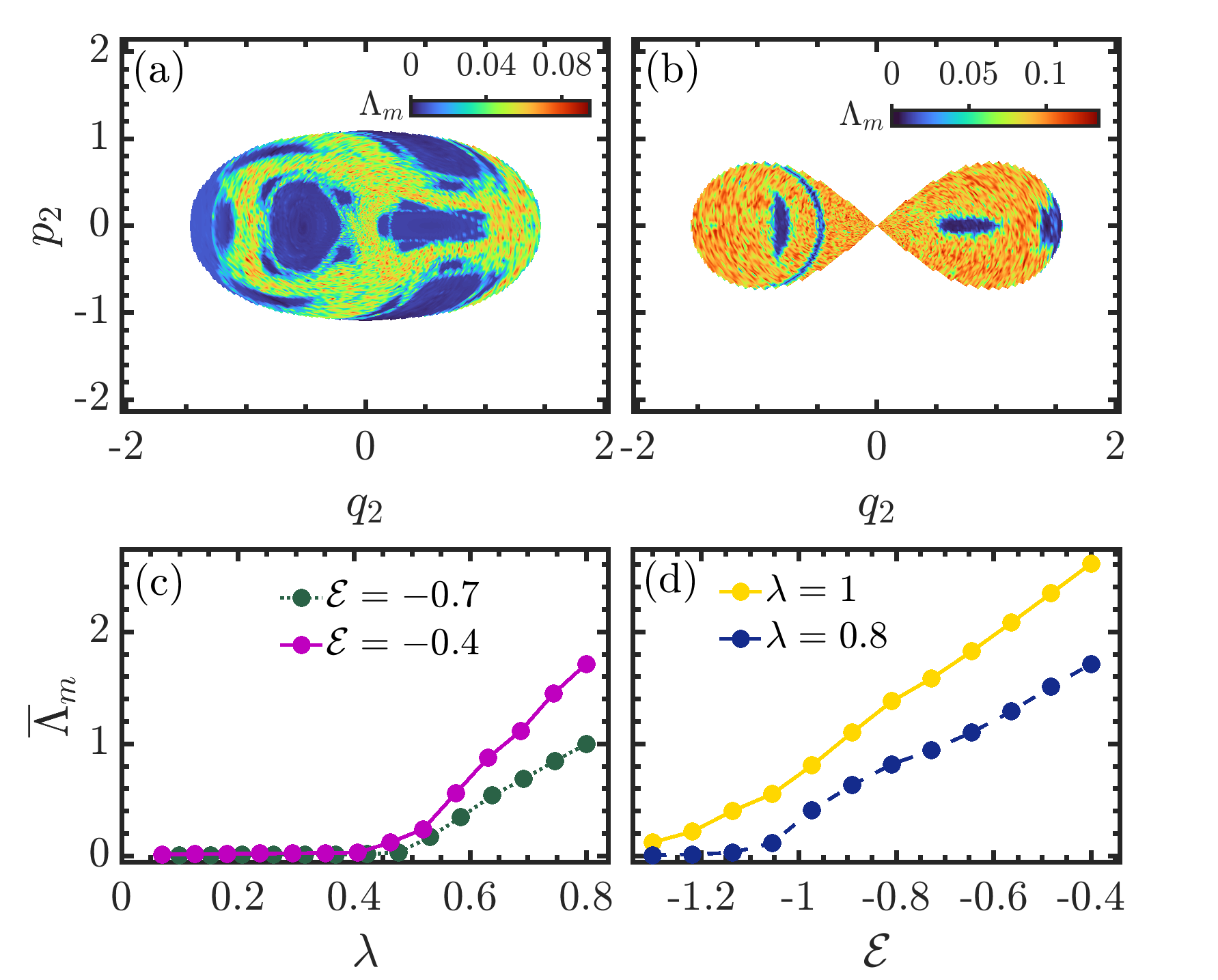}
  \caption{Maximal Lyapunov exponent, $\Lambda_m$, as a function of $q_2$ and $p_2$ for
  (a) $\lambda=0.5, \mathcal{E}=-0.4$ and (b) $\lambda=0.8, \mathcal{E}=-1$.
  The maximal Lyapunov exponent is calculated on a grid of $150\times150$ initial conditions,
  each has evolved up to time $t=1000$.  
  (c) Phase space averaged maximal Lyapunov exponent, $\overline{\Lambda}_m$, as a function 
  of $\lambda$ for different energies.
  (d) $\overline{\Lambda}_m$ as a function of energy for different coupling strengths $\lambda$.
  Other parameters: $\omega=\omega_0=1$.
  All quantities are dimensionless.}
  \label{FgLyp}
 \end{figure}

It is known that the classical Dicke model undergoes a transition from integrability to chaos 
with increasing either the coupling strength with fixed energy or 
the energy at fixed $\lambda$ which satisfies $\lambda>\lambda_c$. 
To see this, we exploit the Poincar\'{e} sections to qualitatively demonstrate 
how chaos emerges as the coupling strength or energy varies. 
Here, we define the Poincar\'{e} section for a given energy $\mathcal{E}$ as the 
intersection of the classical trajectories with the surface that is the plane 
in the variables $(p_2,q_2)$ with $p_1=0$ and $q_1$ being determined by
the energy conservation condition $\mathcal{H}_c=\mathcal{E}$.
By solving the quadratic equation $\mathcal{H}_c(q_1,q_2,p_1=0,p_2)=\mathcal{E}$, 
we get two different values of $q_1$:
\be\label{Egsf}
  q_{1,\pm}=-\frac{\lambda}{\omega} q_2\sqrt{4-p_2^2-q_2^2}\pm
         \sqrt{\frac{\lambda^2}{\omega^2}q_2^2(4-p_2^2-q_2^2)-\omega_0(p_2^2+q_2^2)+2(\omega_0+\mathcal{E})}.
\ee
Moreover, we only record the traversals with $q_1>0$.  

The Poincar\'{e} sections for several coupling strengths with $\mathcal{E}=-0.4$ are 
plotted in Figs.~\ref{CPS}(a)-\ref{CPS}(c). 
At weak coupling strength, as seen in Fig.~\ref{CPS}(a), the Poincar\'{e} section 
exhibits a regular structure, indicating that the system is governed by the regular dynamics.
The chaotic orbits appear as the coupling strength increases, 
as exemplified in Fig.~\ref{CPS}(b) for $\lambda=0.5$ case, 
where the Poincar\'{e} section has complex structure with 
regular islands embedded in the chaotic sea. 
When the coupling strength is sufficiently strong, the system 
becomes a fully chaotic system whose
Poincar\'{e} section is erratically covered by all orbits from 
a set of disordered points, as illustrated in Fig.~\ref{CPS}(c).
The dependence of the degree of chaos on the system 
energy is displayed in Figs.~\ref{CPS}(d)-\ref{CPS}(f),
where we show the Poincar\'{e} sections for different 
system energies with $\lambda=0.8$.
A transition of the Poincar\'{e} section from the regular pattern 
to a globally chaotic sea is clearly visible. 
Hence, the level of chaos in the Dicke model depends on 
both the coupling strength and the system energy. 

To quantitatively analyze the development of chaos depending upon the 
strength of coupling $\lambda$ and the system energy $\mathcal{E}$, 
we consider the maximal Lyapunov exponent, which
measures the rate of exponential separation between two infinitesimally closed orbits.
For the Dicke model, it can be calculated as \cite{Carlos2016,Skokos2010}
\be
  \Lambda_m=\lim_{t\to\infty}\lim_{||\delta{\bf{x}}_0||\to0}\frac{1}{t}
     \ln\frac{||\Omega(t)\cdot{\delta\bf{x}}_0||}{||{\delta\bf{x}}_0||},
\ee 
where ${\delta\bf{x}}_0=(\delta q_{1,0},\delta q_{2,0},\delta p_{1,0},\delta p_{2,0})^\mathrm{T}$ 
is the initial deviation between two given orbits and $\Omega(t)$ 
is the fundamental matrix \cite{Skokos2010}, determined by the equation
\be 
   \dot{\Omega}(t)=\left[\mathbf{J}_4\cdot D^2\mathcal{H}_c({\bf{x}}(t))\right]\cdot\Omega(t).
\ee 
Here, the initial condition is $\Omega(0)=\mathbb{I}_4$ with $\mathbb{I}_4$ 
being the $4\times4$ identity matrix, while 
$\mathbf{J}_4$ and $D^2\mathcal{H}_c({\bf{x}}(t))$ are $4\times 4$ matrices and given by 
\be
  \mathbf{J}_4=
  \begin{pmatrix}
  \mathbf{0}_2  & \mathbb{I}_2 \\
  -\mathbb{I}_2  & \mathbf{0}_2
  \end{pmatrix},\quad
  \left[D^2\mathcal{H}_c(\mathbf{x}(t))\right]_{i,j}
    =\left.\frac{\partial^2\mathcal{H}_c}{\partial x_i\partial x_j}\right|_{\mathbf{x}(t)},\ i,j=1,2,\ldots,4,
\ee
with $\mathbf{0}_2$ is the $2\times2$ zero matrix and $\mathbf{x}(t)$ 
being the solution of the classical equations of motion (\ref{CEM}) at time $t$.

The maximal Lyapunov exponent as a function of $p_2$ and $q_2$ for two different 
combinations of energy $\mathcal{E}$ and coupling strength $\lambda$ are 
plotted in Figs.~\ref{FgLyp}(a) and \ref{FgLyp}(b), respectively.
A comparison with the Poinca\'{e} sections in Figs.~\ref{CPS}(b) and \ref{CPS}(e) 
unveils that the regular and chaotic regions are clearly 
distinguished by $\Lambda_m\approx0$ and $\Lambda_m>0$, as expected.
A further observation of the exhibited behavior of $\Lambda_m$ is that 
it increases as the energy and coupling strength are increased. 

The dependence of the level of chaos on the energy and coupling strength can be
quantitatively captured by the phase space averaged maximal Lyapunov exponent, 
which in this case is also known as the Kolmogorov-Sinai entropy \cite{Lichtenberg2013} 
and defined as
\be
  \overline{\Lambda}_m=\int d\mathcal{S}\Lambda_m,
\ee
where $d\mathcal{S}=rdrd\theta$ is the phase space area element in polar coordinates.
In Fig.~\ref{FgLyp}(c), the variation of $\overline{\Lambda}_m$ with coupling strength $\lambda$
is shown for different energies.
For the system with $\lambda\lesssim0.4$, $\overline{\Lambda}_m=0$, regardless of the energy,
suggesting the regular dynamics of the system. 
As soon as $\lambda\gtrsim0.5$, $\overline{\Lambda}_m$ starts to 
increase with increasing $\lambda$ independent of the energy,
indicating the development of chaos in the system.
It is remarkable that this happens approximately at the 
critical point, $\lambda_c$, of the quantum phase transition. 
Additionally, it is obvious that the larger the energy, the 
faster the growth of $\overline{\Lambda}_m$.
Besides, with increased system energy
the onset of chaos happens at smaller value of the coupling strength.  
These observations indicate that the system energy 
strongly affects the degree of chaos.
The impact of energy on the level of chaos is clearly unveiled in Fig.~\ref{FgLyp}(d), 
where we plot $\overline{\Lambda}_m$ as a function of energy 
$\mathcal{E}$ for different values of $\lambda$.
One can see that $\overline{\Lambda}_m$ exhibits an obvious transition 
from very tiny values to large values with increasing energy.
We also note that the transition to
chaos shifts to lower energy with increasing the strength of coupling. 
It is worth mentioning that the above features of $\overline{\Lambda}_m$ are in consistent with 
the Poincar\'{e} sections shown in Fig.~\ref{CPS}.

 \begin{figure}
  \includegraphics[width=\columnwidth]{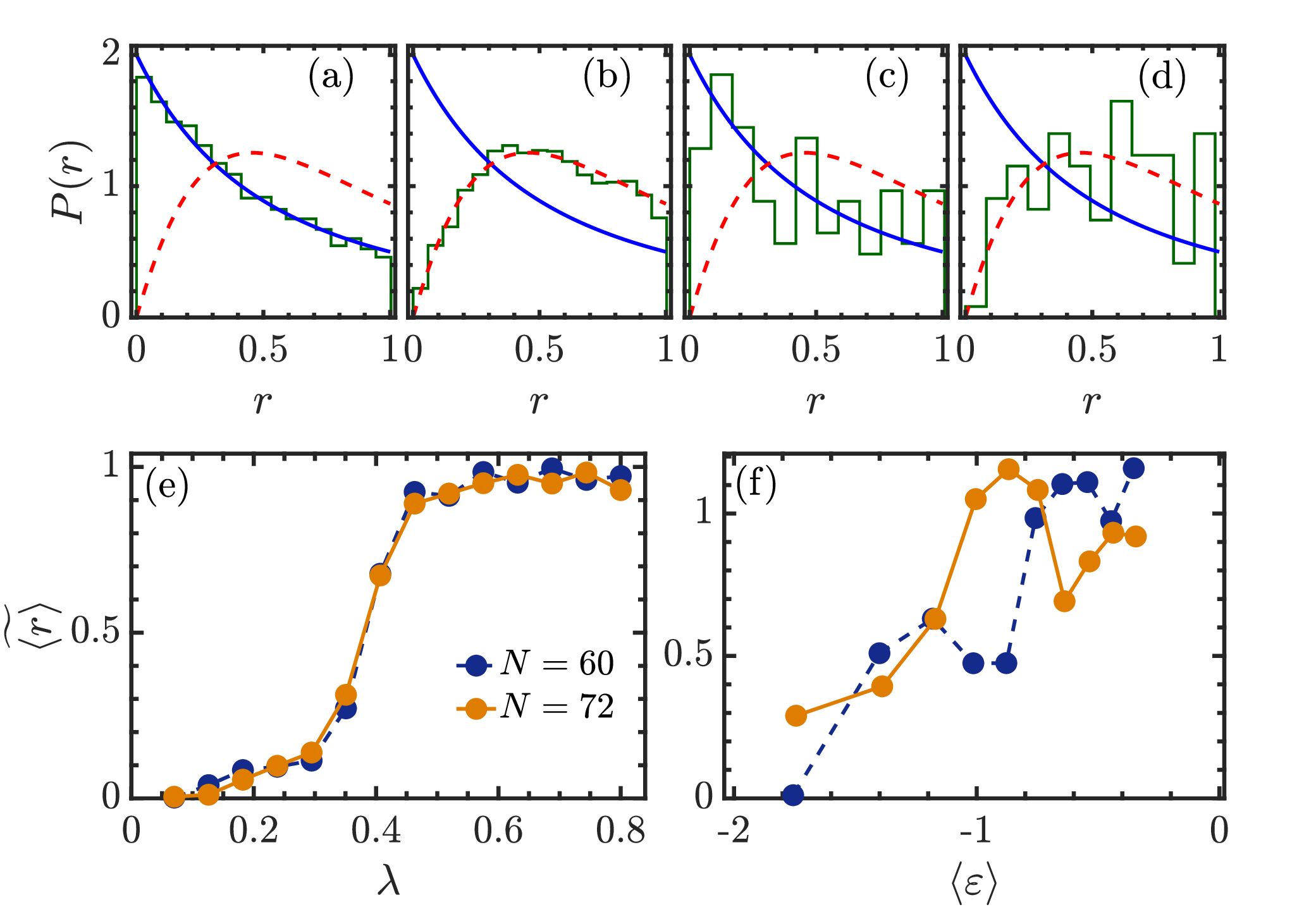}
  \caption{(a)-(b) Level spacing ratio distribution $P(r)$ for (a) $\lambda=0.1$ 
  and (b) $\lambda=1$ with $N=2j=60$.
  (c)-(d) $P(r)$ for different energy intervals: (c) $\varepsilon_n\in[\varepsilon_0,\varepsilon_{149}]$ 
  and (d) $\varepsilon_n\in[\varepsilon_{244},\varepsilon_{393}]$,
  with $N=2j=60$ and $\lambda=1$.
  Here, $\varepsilon_n=E_n/j$ is the $n$th rescaled eigenenergy with $n=0$ corresponding to the ground state.
  The red dashed and blue solid curves in panels (a)-(d) are $P_{GOE}$ and $P_{P}(r)$, respectively.
  (e) Rescaled average level spacing ratio, $\widetilde{\la r\ra}$, 
  as a function of coupling strength $\lambda$ for several system sizes. 
  (f) $\widetilde{\la r\ra}$ as a function of $\la\varepsilon\ra$ 
  for the system sizes as in panel (e) with $\lambda=1$.
  Here, $\la\varepsilon\ra=\sum\varepsilon_n/150$ is the 
  averaged energy of each energy intervals.
  Other parameters: $\omega=\omega_0=1$. 
  The bosonic mode has been truncated at $\mathcal{N}_{trc}=300$.
  All quantities are dimensionless.}
  \label{Dispr}
 \end{figure}

\subsection{Chaos in quantum system} \label{SecB}

Let us now turn to chaotic transition in the quantum model (\ref{DickeH}).
The presence of quantum chaos can be diagnosed by numerous indicators, including
spectral and eigenstates statistics 
\cite{Bohigas1984,Izrailev1990,Zyczkowski1990,Gomez2011,Haake2019,Borgonovi2016}, 
various dynamical probes \cite{Borgonovi2016,Gorin2006,Arpan2022},
and different complexities \cite{Parker2019,Dymarsky2020,Arpan2021a,Arpan2021b,Vijay2022,Espanol2023,Hashimoto2023}, 
to mention a few.
Here, we focus on the distribution of level spacing ratios, defined as \cite{Oganesyan2007,Atas2013}
\be
  r_n=\mathrm{min}\left(\delta_n,\frac{1}{\delta_n}\right),
\ee
where $\delta_n=s_n/s_{n+1}$ with $s_n=E_{n+1}-E_n$ being the consecutive level spacing 
for ordered energy levels $E_n<E_{n+1}$ of the Hamiltonian (\ref{DickeH}). 
As investigating spacing ratio distribution is free from the 
need of performing the unfolding procedure for energy spectrum, 
it turns out to be the most widely used chaos indicator 
for studying quantum many-body chaos 
\cite{Corps2020,Lucas2020,Patrycja2022,Giraud2022,Pausch2021,Nakerst2023}. 
Another advantage of using $r_n$ instead of $s_n$ as chaos indicator is that 
the support of the distribution of $r_n$ is bounded, that is $r_n\in[0,1]$.

The distribution of $r_n$ for chaotic systems that belong to the Gaussian
orthogonal ensemble (GOE) is given by \cite{Atas2013,Nakerst2023},
\be
  P_{GOE}(r)=\frac{2}{Z_{GOE}}\frac{r+r^2}{(1+r+r^2)^{5/2}},
\ee 
where $Z_{GOE}=8/27$ is the normalization constant.
For integrable systems, uncorrelated eigenvalues exhibiting 
Poisson statistics, we have \cite{Atas2013}
\be
  P_P(r)=\frac{2}{(1+r)^2}.
\ee
A prominent feature of the chaotic spectra is level repulsion characterized by $P_{GOE}(0)=0$,
in contrast to Poisson spectra with $P_P(0)\neq0$.

In Figs.~\ref{Dispr}(a) and \ref{Dispr}(b), we show the ratio distribution $P(r)$ for two
different values of $\lambda$ with system size $N=60$. 
The distribution $P(r)$ in Fig.~\ref{Dispr}(a) for small $\lambda$ is well 
captured by the Poisson case, in agreement with the classical 
regular dynamics exhibited in Fig.~\ref{CPS}(a).
The ratio distribution $P(r)$ for $\lambda=1$ in Fig.~\ref{Dispr}(b) demonstrates
an opposite situation, where the GOE distribution of $r$ is clearly visible.
We see again that the behavior of $P(r)$ at $\lambda=1$ is consistent with 
the classical Poincar\'{e} section in Fig.~\ref{CPS}(c). 
To reveal the dependence of chaos on the system energy, 
we consider the distribution $P(r)$ extracted from energy levels 
within a certain energy interval, as has been done in Refs.~\cite{Pausch2021,Nakerst2023}. 
The energy interval in our study is an interval between 
$\varepsilon_n$ and $\varepsilon_{n+149}$ with $\varepsilon_n=E_n/j$ 
being the $n$th rescaled eigenenergy.
We plot $P(r)$ for two representative energy intervals 
$\varepsilon_n\in[\varepsilon_0,\varepsilon_{149}]$ and
$\varepsilon_n\in[\varepsilon_{244},\varepsilon_{393}]$ 
in Figs.~\ref{Dispr}(c) and \ref{Dispr}(d), respectively. 
Here, $\varepsilon_0$ denotes the rescaled ground state energy.
It can be clearly seen that the ratio distribution $P(r)$ is close 
to the Poisson distribution $P_P(r)$ for the lowest energy levels, 
while it turns to follow the GOE distribution $P_{GOE}(r)$ 
for the energy levels at high energy.
In particular, $P(r=0)$ exhibits an obvious transition from
$P(r=0)\neq0$ to $P(r=0)\approx0$ as the energy of energy interval is increased.

 \begin{figure}
  \includegraphics[width=\columnwidth]{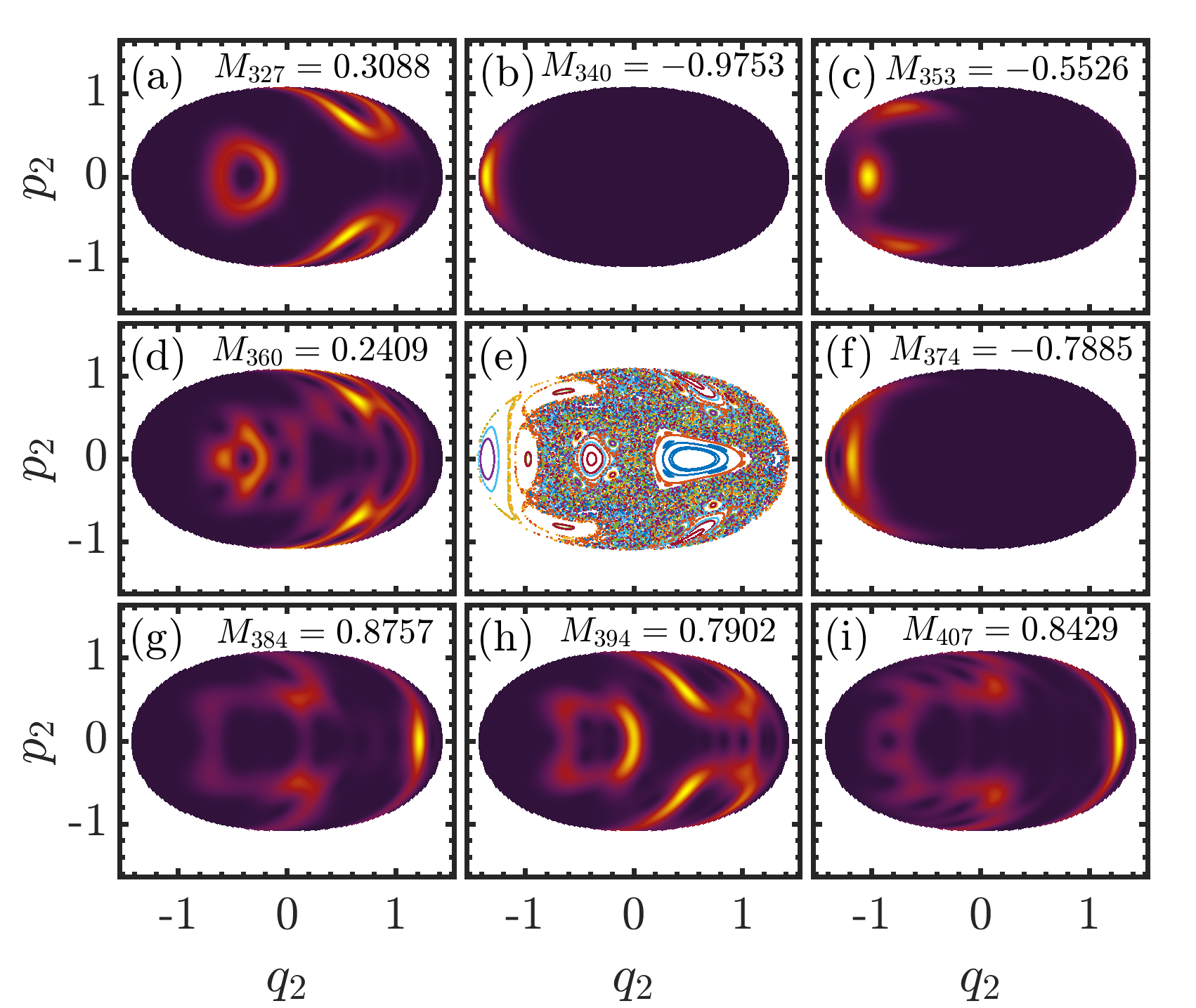}
  \caption{Poincar\'{e}-Husimi function, $Q_n^P(p_2,q_2)$ 
  for several positive parity eigenstates with energies: (a) $E_{327}/j=-0.4391$,
  (b) $E_{340}/j=-0.4297$, (c) $E_{353}/j=-0.414$, (d) $E_{360}/j=-0.4067$,
  (f) $E_{374}/j=-0.3955$, (g) $E_{384}/j=-0.3856$, (h) $E_{394}/j=-0.3754$,
  and (i) $E_{407}/j=-0.3641$.
  Phase space overlap index $M_n$ obtained from Eq.~(\ref{Mdex}), corresponding to
  these eigenstates are mentioned in the figure. 
  The classical Poincar\'{e} section for $\mathcal{E}=-0.4$ is plotted in the center panel (e).
  Other parameters: $\omega=\omega_0=1$, $N=2j=100$ and $\lambda=0.47$. 
  The bosonic mode has been truncated at $\mathcal{N}_{trc}=170$.
  All quantities are dimensionless.}
  \label{FHsf}
 \end{figure}

A more popular quantity that has been employed to 
detect the presence of quantum chaos is
the mean spacing ratio, defined as $\la r\ra=\int_0^1 rP(r)dr$. 
It was known that $\la r\ra$ interpolates between two extreme values 
$\la r\ra_P=2\ln2-1\approx0.38629$ and $\la r\ra_{GOE}=4-2\sqrt{3}\approx0.53590$, 
corresponding to Poisson and GOE distributions, respectively \cite{Atas2013}. 
This leads to the definition of rescaled mean spacing ratio \cite{Patrycja2022}
\be
  \widetilde{\la r\ra}=\frac{|\la r\ra-\la r\ra_P|}{\la r\ra_{GOE}-\la r\ra_P}.
\ee
It varies in the interval $\widetilde{\la r\ra}\in[0,1]$. 
For integrable systems with Poisson distribution, we have $\widetilde{\la r\ra}=0$,    
while chaotic systems with GOE distribution give rise to $\widetilde{\la r\ra}=1$. 

The variation of $\widetilde{\la r\ra}$ as a function of coupling strength 
$\lambda$ for several system sizes is plotted in Fig.~\ref{Dispr}(e).
We see that the overall behavior of $\widetilde{\la r\ra}$ as a function of $\lambda$
is almost unaffected by varying the system size.
Regardless of the system size, $\widetilde{\la r\ra}$
exhibits a rapid growth from small values to the values 
around one in the region $\lambda\in[0.35,0.55]$. 
Again, it is remarkable that this is close to the critical coupling 
strength, $\lambda_c$, of the quantum phase transition.
Thus, the quantum system undergoes a crossover from integrability to chaos
with increasing the strength of coupling, in agreement with the 
behavior of $P(r)$ observed in Figs.~\ref{Dispr}(a) and \ref{Dispr}(b).
On the other hand, the dependence of $\widetilde{\la r\ra}$ on the system energy is
shown in Fig.~\ref{Dispr}(f), where we display how $\widetilde{\la r\ra}$ 
evolves for different energy intervals and system sizes with fixed $\lambda$.
The overall increasing behavior of $\widetilde{\la r\ra}$ as the energy 
interval moves to high energy levels confirms that the degree of chaos in the Dicke model
can be enhanced by increasing the system energy.
As a final remark of this section, we would like to point out that the
behavior of $\widetilde{\la r\ra}$ coincides with $\overline{\Lambda}_m$, 
suggesting a good quantum-classical correspondence.

\section{Overlap index in classical phase space} \label{Third}
  
The aim of this work is to analyze how the mixed eigenstates evolve 
as the semiclassical limit $N\to\infty$ is approached. 
This requires to focus on the cases that have mixed-type 
phase phase in the semiclassical dynamics.
Thus, we restrict our study in the parameter region
$\lambda\in[0.45,0.5]$ with energy $\mathcal{E}\in[-0.7,-0.5]$.
A careful numerical check has verified that the classical dynamics 
for our considered parameters and energy regions 
is indeed characterized by mixed-type phase space.  
Moreover, following our previous works \cite{Batistic2013,Robnik2016,Lozej2022,QianW2023b}, 
we identify the types of eigenstates through the phase space overlap index.
To this end, let us first introduce the definition of Husimi function 
for an individual eigenstate.

\subsection{Husimi function} \label{TrA}

As the Gaussian smoothing of the well-known Wigner function \cite{Wigner1932}, 
the Husimi function \cite{Husimi1940} constitutes the simplest quasiprobability distribution 
of a quantum state in classical phase space and  
has been widely used as a powerful tool to explore 
various properties of quantum eigenstates 
\cite{Cameo2021a,Cameo2021b,Cameo2022,QianW2020,Romera2012,Villasenor2021,
Cibils1992,Oregi2014,Sinha2020,Mondal2020,QianW2021}.  
It is usually defined as the projection of a quantum state onto 
the coherent state \cite{Takahashi1986}.
Thus, the Husimi function for the $n$th eigenstate, $|E_n\ra$, 
of the Dicke model is given by \cite{Aguiar1991,Furuya1992}
\be\label{HusimiF}
   Q_n=|\la\mathrm{CS}|E_n\ra|^2,
\ee
where $|\mathrm{CS}\ra=|\alpha\ra\otimes|\xi\ra$. 
The normalization condition of the coherent states implies that the Husimi function can be normalized as
\be
  \frac{2j+1}{\pi^2}\int_{\mathbb{R}^4}\frac{d^2\alpha d^2\xi}{(1+|\xi|^2)^2}Q_n=1,
\ee
with $j=N/2$ and $d^2w=d\mathrm{Re}(w)d\mathrm{Im}(w)\ (w=\alpha, \xi)$.

 \begin{figure}
  \includegraphics[width=\columnwidth]{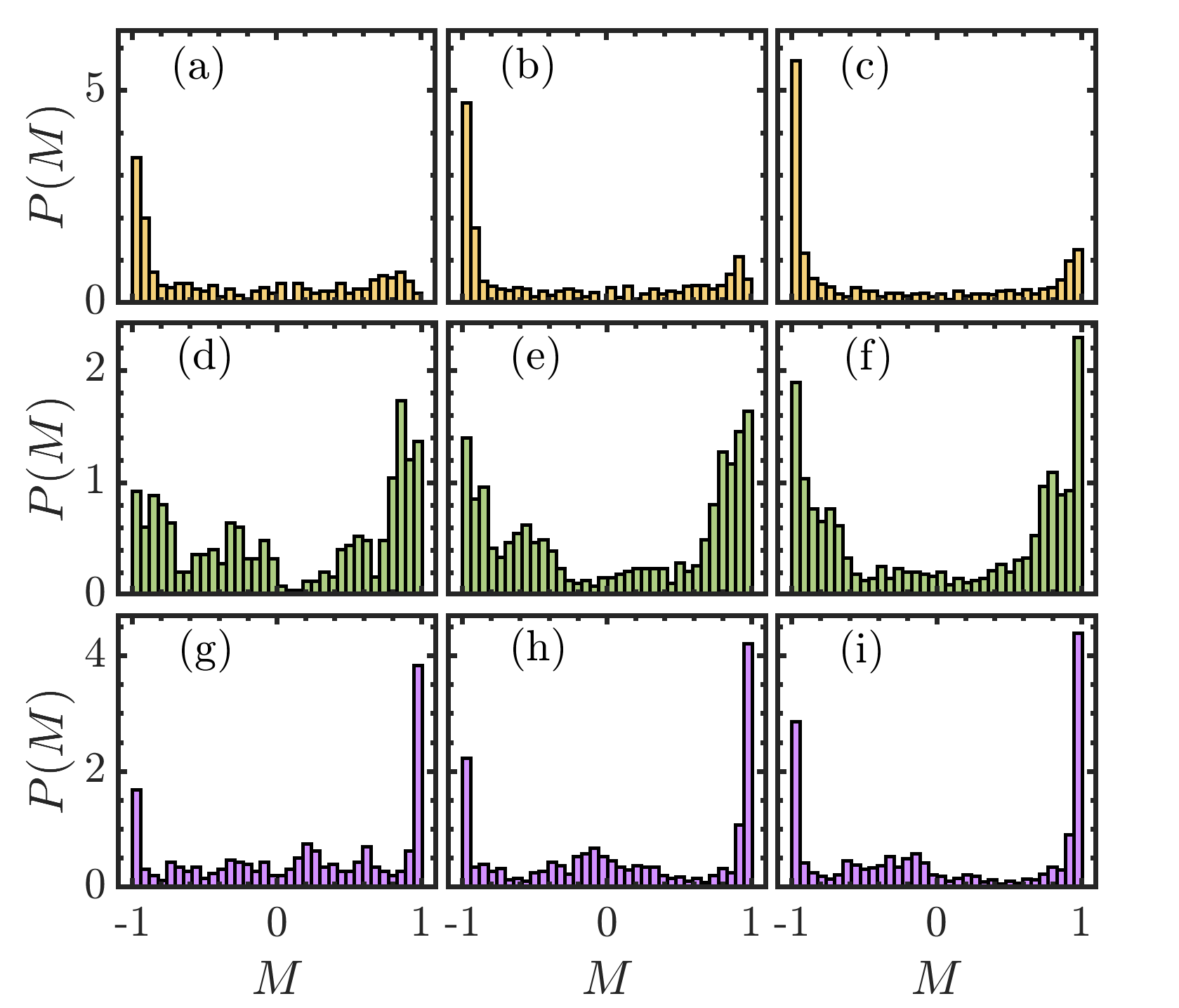}
  \caption{(a)-(c): Probability distribution of the phase space overlap index, $P(M)$,
  for system size ensembles (a) $N\in[72,88]$, (b) $N\in[92,108]$, and (c) $N\in[112,128]$ with
  $\lambda=0.47$ and $\mathcal{E}=-0.5$.
  (d)-(f): $P(M)$ for the same system size ensembles and coupling strength $\lambda$ 
  as in panels (a)-(c), but for the case of $\mathcal{E}=-0.4$.  
  (g)-(i): $P(M)$ for the same system size ensembles as in panels (a)-(c), 
  but with $\lambda=0.5$ and $\mathcal{E}=-0.4$.
  For each case, the energy of the considered eigenstates satisfies 
  $E_n/j\in[\mathcal{E}-\delta E,\mathcal{E}+2\delta E]$ with $\delta E=0.04$.  
  Other parameters: $\omega=\omega_0=1$. 
  The bosonic mode has been truncated at $\mathcal{N}_{trc}=170$.
  For each ensemble, the system size $N$ is increased in step of $4$.
  All quantities are dimensionless.}
  \label{MdxPrb}
 \end{figure}

For Hamiltonians with two degrees of freedom, such as the Dicke model,
the Husimi function $Q_n$ in (\ref{HusimiF}) has four real variables.
As a consequence, it is difficult to visualize and display it.
A common way to circumvent this difficulty is 
to study different projections or sections of the Husimi function 
\cite{Bakemeier2013,Leboeuf1990,Groh1998,Magnani2016}.  
For our purpose, we consider the Poincar\'{e}-Husimi function 
which is evaluated along the Poincar\'{e} surface $q_{1,+}$ 
defined in Eq.~(\ref{Egsf}) and calculated as
\be
  Q_n^P(p_2,q_2)=|\la p_1=0,q_1=q_{1,+};p_2,q_2|E_n\ra|^2.
\ee
Here, the energies of eigenstates are chosen to satisfy 
$|E_n/j-\mathcal{E}|\simeq0$ for a certain value of $\mathcal{E}$,
so that they correspond to a well defined classical dynamical regime (phase portrait). 
Now, $Q_n^P(p_2,q_2)$ is a function of two variables and its density plot provides 
a quite illustrative comparison to the classical Poincar\'{e} section \cite{Bakemeier2013,Magnani2016}.

The Poincar\'{e}-Husimi function for several eigenstates with 
energy close to $\mathcal{E}=-0.4$ are plotted in Fig.~\ref{FHsf}.
Comparing with the classical Poincar\'{e} section, which is shown in Fig.~\ref{FHsf}(e),
we see that the eigenstates can be qualitatively divided into three types. 
The first type corresponds to the eigenstates whose Poincar\'{e}-Husimi functions
are mainly located in the classical regular regions, as evident in Figs.~\ref{FHsf}(b) and \ref{FHsf}(f). 
This type of eigenstates is usually called regular eigenstates.
In contrast to the regular eigenstates, the Poincar\'{e}-Husimi functions 
in Figs.~\ref{FHsf}(g)-\ref{FHsf}(i) are almost fully distributed over the chaotic sea.
This allows us to name this type of eigenstates the chaotic eigenstates.
Apart from these two extreme cases, we still have some eigenstates, 
referred to as the mixed eigenstates, 
whose Poincar\'{e}-Husimi functions occupy both regular and chaotic regions, 
as exemplified in Figs.~\ref{FHsf}(a), \ref{FHsf}(c), and \ref{FHsf}(d). 
The presence of the mixed eigenstates originates from different tunneling 
processes between various structures in classical phase space. 

A natural question about the existence of different types of eigenstates 
is how to quantitatively characterize them. 
In particular, unveiling how the relative proportion of the mixed eigenstates varies 
as the system approaches the semiclassical limit would 
improve our understanding of generic quantum systems whose 
classical counterpart has mixed-type phase space. 
For single particle systems, our previous works \cite{Lozej2022,QianW2023b} 
have shown that the relative proportion of 
the mixed eigenstates undergoes a power-law decay as
the semiclassical limit is approached.
But, whether this conclusion still holds in many-body quantum systems remains unknown.
In the following of this section, we address above mentioned questions in the Dicke model.

\subsection{Phase space overlap index} \label{TrB}

As has been done in Refs.~\cite{Batistic2013,Robnik2016,Lozej2022,QianW2023b}, 
we discriminate the types of the eigenstates by means of 
the phase space overlap index, $M$, defined as
the overlap of Poincar\'{e}-Husimi function 
with different regions of phase space.
Specifically, by employing the polar coordinates $(r,\theta)$ with $0\leq r\leq2$,
we separate the classical phase space $(q_2,p_2)$ into
a grid with $\mathcal{N}$ small cells, which are labeled by $(i,j)$ with
$q_2(i,j)=r_i\cos\theta_j$ and $p_2(i,j)=r_i\sin\theta_j$. 
Consequently, the $n$th eigenstate Poincar\'{e}-Husimi function is 
now discretized as $Q_n^P(i,j)=Q_n^P[p_2(i,j),q_2(i,j)]$
with normalization condition $\sum_{i,j}^\mathcal{N}Q_n^P(i,j)/\mathcal{N}=1$. 
Then, we set a quantity $C_{i,j}=+1$ to the cells that belong to the chaotic region 
and $C_{i,j}=-1$ for other cells.
Finally, the phase space index for the $n$th eigenstate is defined as
\be\label{Mdex}
  M_n=\frac{1}{\mathcal{N}}\sum_{i,j}Q_n^P(i,j)C_{i,j}.
\ee
The definition of $M_n$ implies that it varies in the interval $M_n\in[-1,1]$ with 
$M_n=+1$ and $-1$ corresponding to fully chaotic and regular eigenstates, respectively. 
For the mixed eigenstates, as their Poincar\'{e}-Husimi functions are distributed in both
chaotic and regular regions, we have $-1<M_n<1$. 
The values of $M_n$ for the eigenstates in Fig.~\ref{FHsf} are given in the figures.
One can see that $M_n$ is either close to $+1$ or $-1$ for chaotic and regular eigenstates,
while it takes the values between $-1$ and $+1$ for mixed eigenstates, as expected.

 \begin{figure}
  \includegraphics[width=\columnwidth]{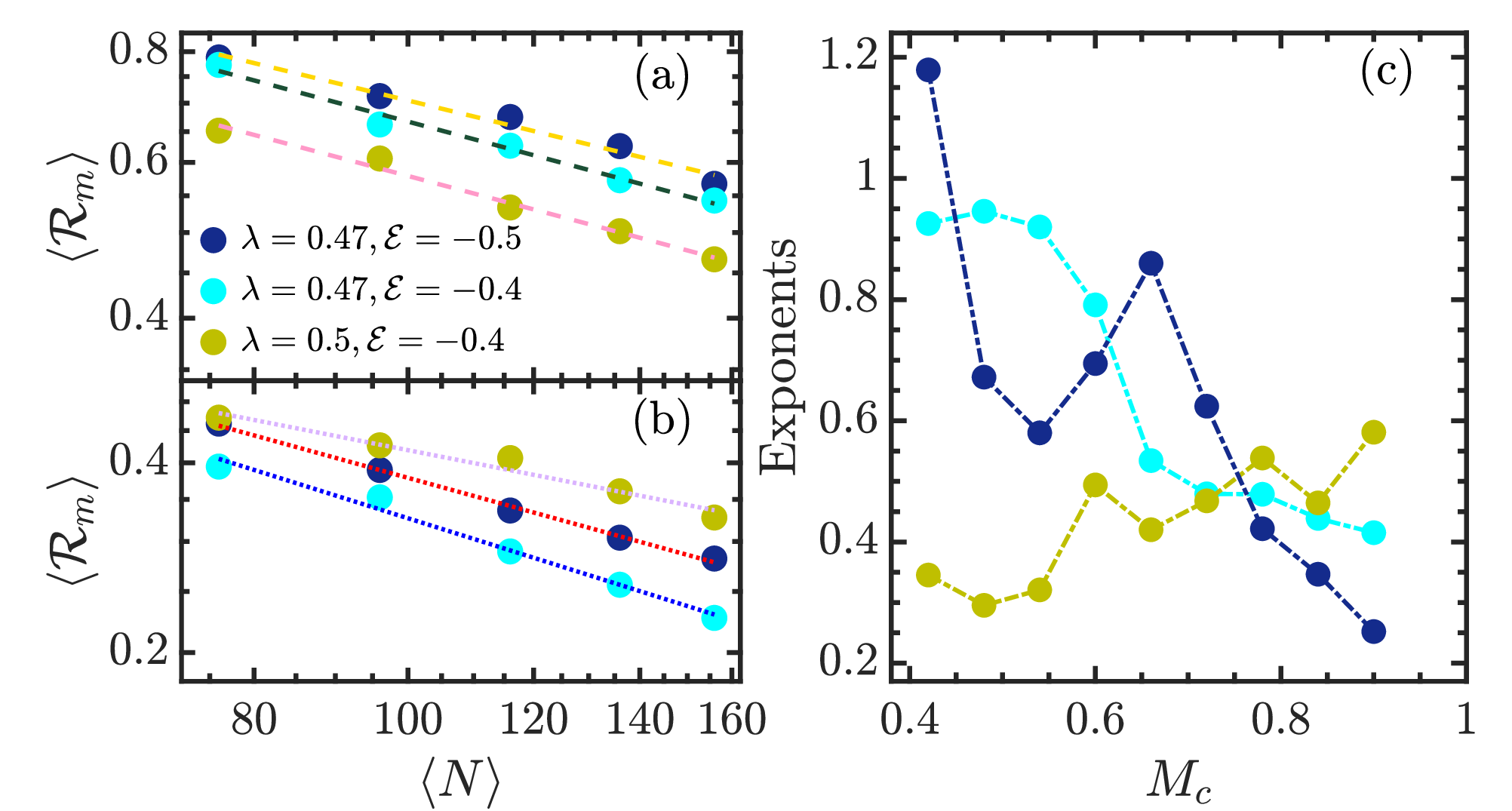}
  \caption{(a): Ensemble averaged $\mathcal{R}_m$ [cf.~Eq.~(\ref{RelativeR})] as a function
  of ensemble averaged system size $\la N\ra$ for $M_c=0.8$ and 
  several combinations of coupling strength $\lambda$ and system energy $\mathcal{E}$.
  The dashed lines are the power law behavior of the form $\la\mathcal{R}_m\ra\propto\la N\ra^{-\gamma}$
  with $\gamma=0.4357, 0.4797$, and $0.4779$ (from top to bottom).
  (b) $\la\mathcal{R}_m\ra$ as a function of $\la N\ra$ for the same combinations of $\lambda$ and $\mathcal{E}$
  as in panel (a) and $M_c=0.6$. 
  Dotted lines denote the power-law $\la\mathcal{R}_m\ra\propto\la N\ra^{-\gamma}$ with
  $\gamma=0.4941, 0.6944$ and $0.7913$ (from top to bottom).
  (c): Exponents of the power-law decay as a function of $M_c$ for the same combinations of 
  $\lambda$ and $\mathcal{E}$ as in panels (a) and (b). 
  For each case, the energy of the considered eigenstates satisfies 
  $E_n/j\in[\mathcal{E}-\delta E,\mathcal{E}+2\delta E]$ with $\delta E=0.04$. 
  Here, the system ensemble for each case is the same as Fig.~\ref{MdxPrb} and
  the system size in increased in step $4$ in each ensemble.    
  Other parameters: $\omega=\omega_0=1$. 
  The bosonic mode has been truncated at $\mathcal{N}_{trc}=170$.
  All quantities are dimensionless.}
  \label{Rmix}
 \end{figure}

More information about the properties of the eigenstates is revealed 
by the probability distribution of $M_n$, defined as
\be
   P(M)=\int\rho(M)dM_n,
\ee
where $\rho(M)=\sum_n\delta(M-M_n)$ is the probability density function. 
It was known that $P(M)$ has two peaks at $M=\pm1$ 
for single particle systems \cite{Lozej2022,QianW2023b}, 
but it is necessary to perform a detailed investigation in quantum many-body systems,
like the Dicke model, to see whether this property of $P(M)$ still holds.
It is worth pointing out that the structure of the classical Poincar\'{e} section only depends on 
the coupling strength and energy.    
This allows us to get sufficient data for the statistics of $M$ by considering 
eigenstates within an energy window for an ensemble of different system sizes.
In this work, for a given energy $\mathcal{E}$, we will focus on the 
eigenstates whose energies satisfy 
$E_n/j\in[\mathcal{E}-\delta E,\mathcal{E}+2\delta E]$ with $\delta E=0.04$. 
We have checked that the underlying classical dynamics keeps unchanged 
for our considered energy interval. 

In Fig.~\ref{MdxPrb}, the distribution $P(M)$ for different system ensembles 
and coupling strengths, as well as energies are depicted.
$P(M)$ is calculated from a finite number of eigenstates with energies 
within a given energy window, and plotted as the histogram.  
However, one can expect that it will converge to a smooth distribution 
in the classical limit $N\to\infty$.   
As seen from the figures, although $P(M)$ exhibits large fluctuations 
compared to the cases of single particle systems,
it is still behaved as a double peak distribution 
with two peaks located at $M\simeq\pm1$, respectively.
This is more evident for the large system sizes, 
as shown in Figs.~\ref{MdxPrb}(c), \ref{MdxPrb}(f), and \ref{MdxPrb}(i).
The large fluctuations observed in the behaviors of $P(M)$
are due to the relative small system size, which leads to the 
wide support of the Poincar\'{e}-Husimi function in the phase space.
We further observe that the fluctuations in $P(M)$
can be suppressed by increasing either the system 
size or the degree of chaos, as revealed in single particle systems \cite{Lozej2022,QianW2023b}.  

The above observations of $P(M)$ suggest that 
the larger is the system size $N$, the sharper 
the double peak shape of $P(M)$.
Hence, one can expect that only the regular and chaotic 
eigenstates are left in the semiclassical limit $N\to\infty$.
To strengthen this statement, we study how the relative proportion 
of the mixed eigenstates changes as the system size is increased.
In our consideration, the mixed eigenstates are identified as 
the eigenstates that satisfy $-M_c\leq M\leq M_c$ with the cutoff $M_c=0.8$. 
There is an arbitrariness in the choice of the value of $M_c$,
however, our main results are unaffected by the exact cutoff value of $M_c$.

The relative proportion of the mixed eigenstates is defined as
\be \label{RelativeR}
  \mathcal{R}_m=\frac{\Delta\mathcal{N}_{mix}}{N_{\delta E}},
\ee 
where $N_{\delta E}=\sum_{E_n/j\in\delta E}$ denotes the number of eigenstates in 
above mentioned energy window, while $\Delta\mathcal{N}_{mix}=\sum_{M_n\in[-M_c,M_c]}$ 
represents the number of mixed eigenstates in the same energy window. 
In Figs.~\ref{Rmix}(a) and \ref{Rmix}(b), we show how the ensemble 
averaged relative proportion of the mixed eigenstates, denoted by $\la\mathcal{R}_m\ra$,
varies as a function of ensemble averaged system size $\la N\ra$ for 
different values of $M_c$ and several combinations of $\lambda$ and $\mathcal{E}$.
As we see in the figures, the relative proportion of the mixed 
eigenstates decreases with increasing system size, regardless 
of the values of $\lambda$ and $\mathcal{E}$, as well as $M_c$. 
In particular, we find that the functional relationship between $\la\mathcal{R}_m\ra$ and $\la N\ra$ is well
captured by the power-law decay of the form $\la\mathcal{R}_m \ra\propto\la N\ra^{-\gamma}$ 
with exponent $\gamma$ being a function of the control parameters. 
Similar power-law behavior has also been observed in the single particle systems \cite{Lozej2022,QianW2023b}, 
indicating that the power-law decay of the relative proportion of the mixed eigenstates 
in approaching the semiclassical limit should be a universal behavior.
Moreover, the decreasing of $\mathcal{R}_m$ as the system 
size increases is also consistent with the PUSC, and, thus, provides the 
evidence for the correctness of the PUSC in mixed-type many-body quantum systems.
This means that a mixed-type many-body system has only fully regular 
and chaotic eigenstates in the ultimate semiclassical limit.

We note that the independence of our results upon the choice 
of $M_c$ is confirmed by the overall power-law decay of 
$\mathcal{R}_m$ for different values of $M_c$, 
as illustrated in Figs.~\ref{Rmix}(a) and \ref{Rmix}(b). 
However, from the same figures, we see that the power-law exponent $\gamma$ is 
strongly dependent on the value of $M_c$. 
This is more evident in Fig.~\ref{Rmix}(c), where we display the 
variation of the exponent as a function of  $M_c$ for several 
combinations of $\lambda$ and $\mathcal{E}$.
One can clearly see that the exponent exhibits  quite large fluctuations in its behavior.
This can be attributed to the fluctuations in $P(M)$ that show an 
obvious dependence on the choice of $M_c$ for 
certain control parameters [cf. Fig.~\ref{MdxPrb}]. 
As the $P(M)$ becomes smoother by increasing the level of chaos, 
the fluctuations in the behavior of the exponent should undergo a significant 
suppression for high degree of chaoticity, 
as exemplified in the case of $\lambda=0.5$ and $\mathcal{E}=-0.4$.

\section{Conclusions}\label{Fourth}

A basic and valuable tool for comprehending various characters 
of general mixed quantum systems is the 
principle of uniform semiclassical condensation of Wigner or Husimi functions (PUSC) \cite{Robnik2019,Robnik2020}. 
Although the validity of the PUSC in the mixed-type single particle 
systems has been verified by numerous works, its correctness for the   
mixed-type many-body quantum systems remains unknown.
With the aim to address this question, we have delved into
a detailed analysis of the eigenstates and their properties in the celebrated Dicke model.
As a prototypical model in the studies of both quantum and classical chaos,
the Dicke model is a very suitable many-body quantum model for our study 
because it attains its well defined semiclassical limit as the system size is increased. 
Hence, the Dicke model allows us to define an effective Planck constant, which is 
essential to explore how the features of eigenstates 
change as the semiclassical limit is approached.

By investigating the development of chaos in both classical and quantum Dicke model,
we have determined the parameter and energy regions that exhibit the mixed-type behavior. 
By means of Husimi function, we have shown that the eigenstates can be divided into different types.
To quantitatively characterize different types of eigenstates, 
we have calculated the phase space overlap index, which is defined in terms of the Husimi function
and acts as a useful tool for studying the characters of the eigenstates. 
For the eigenstates that belong to a certain classical dynamical regime, 
the distribution of their phase space overlap index is well described by double peak distribution with
two peaks at $-1$ and $+1$, corresponding to the fully regular and chaotic eigenstates, respectively. 
For finite system size, there are many mixed eigenstates with the value of phase space 
overlap index varying between $-1$ and $+1$.
However, we have shown that the relative proportion of the mixed 
eigenstates vanishes with increasing system size, i.~e. approaching the semiclassical limit.
This means that the eigenstates in the mixed-type many-body quantum systems belong 
to either regular or chaotic type as the semiclassical limit is approached,
similarly as we have observed in the single particle systems \cite{Lozej2022,QianW2023b}
and in accord with the PUSC.
In particular, we have demonstrated that the decay of the relative 
proportion of the mixed eigenstates with increasing system size 
is well captured by a power-law, similar to the results obtained in the single particle systems.  
This leads us to believe that the power-law decay is a universal behavior in 
the evolution of the relative proportion of the mixed eigenstates
in mixed-type systems as the semiclassical limit is approached .

An interesting extension of the present work is to explore the validity of PUSC in
many-body quantum systems that have no classical correspondence, 
such as various quantum spin systems.
The lack of classical counterpart indicates that new methods 
should be found to identify the types of eigenstates.
A possible candidate to fulfill this requirement might be the entanglement entropy of the eigenstates,
however, we leave this subtle question for our future investigation.
In addition, a theoretical understanding of the properties exhibited by the mixed eigenstates 
in many-body quantum systems is also deserved for future exploration. 
Finally, we hope that our work could motivate more studies of the features 
of the mixed-type many-body quantum systems.

\acknowledgments  

This work was supported by the Slovenian Research and Innovation Agency (ARIS) under the 
Grants Nos.~J1-4387 and P1-0306. Q.~W. acknowledges support from the 
National Science Foundation of China (NSFC) under Grant No.~11805165,
Zhejiang Provincial Nature Science Foundation under Grant No.~LY20A05001.

\bibliographystyle{apsrev4-1}
\bibliography{DickeMix}

\end{document}